\documentclass[10pt,journal,compsoc]{IEEEtran}
\usepackage{graphicx}
\usepackage{subcaption}
\usepackage{array}
\usepackage{hyperref}
\usepackage[english]{babel} 
\usepackage[dvipsnames]{xcolor}
\newcommand{\Code}[1]{\textcolor{teal}{#1}}
\usepackage{fvextra} 
\usepackage[most]{tcolorbox}
\DefineVerbatimEnvironment{prompt}{Verbatim}{
    commandchars=\\\{\}, 
    formatcom=\color{black}, 
    breaklines=true, 
    breakanywhere=true, 
    fontsize=\small
}
\tcbset{
  prompt/.style={
    colback=gray!10!white,    
    colframe=gray,           
    rounded corners,         
    boxrule=0.4mm,           
    fonttitle=\bfseries,     
    enhanced,                
    breakable,               
  }
}
\newcommand{\Title}{Blockchain Data Analysis\\ in the Era of Large-Language Models}
\title{\Title}
\author{
    Kentaroh Toyoda, Xiao Wang, Mingzhe Li, Bo Gao, Yuan Wang, and Qingsong Wei
    \thanks{
        Institute of High Performance Computing (IHPC), Agency for Science, Technology and Research (A*STAR), 1 Fusionopolis Way, \#16-16 Connexis, Singapore 138632, Republic of Singapore.\\
        \url{kentaroh.toyoda@ieee.org}
    }
}

\begin{document}


\maketitle

\begin{abstract}
Blockchain data analysis is essential for deriving insights, tracking transactions, identifying patterns, and ensuring the integrity and security of decentralized networks. It plays a key role in various areas, such as fraud detection, regulatory compliance, smart contract auditing, and decentralized finance (DeFi) risk management. However, existing blockchain data analysis tools face challenges, including data scarcity, the lack of generalizability, and the lack of reasoning capability. 

We believe large language models (LLMs) can mitigate these challenges; however, we have not seen papers discussing LLM integration in blockchain data analysis in a comprehensive and systematic way. This paper systematically explores potential techniques and design patterns in LLM-integrated blockchain data analysis. We also outline prospective research opportunities and challenges, emphasizing the need for further exploration in this promising field. This paper aims to benefit a diverse audience spanning academia, industry, and policy-making, offering valuable insights into the integration of LLMs in blockchain data analysis.
\end{abstract}


\section{Introduction}
Blockchain data analysis involves the examination of data recorded on and off the blockchain networks to derive insights, identify patterns, and monitor activities. 
The need for robust blockchain data analysis emerged alongside the rise of cryptocurrencies like Bitcoin. Early use cases focused on tracing illicit transactions and addressing concerns about the misuse of pseudonymous networks (e.g., SilkRoad \cite{christin2013SilkRoad}). Today, blockchain data analysis spans diverse domains, from auditing smart contracts and detecting network anomalies to predicting market trends and assessing the impact of governance proposals. 

We have seen a variety of methods that analyze on-chain and off-chain data using statistics, machine learning, graph analysis, and natural language processing (NLP). While traditional analytical methods have provided valuable insights, they often face challenges such as limitations due to the scarcity of ground truth, the lack of generalizability, and the lack of explainability.

We believe that large language models (LLMs) will overcome these challenges and have enormous potential as a core technology for a wide range of blockchain data analysis tasks. In particular, LLMs could move the needle in blockchain data analysis from the following perspectives. 
\begin{enumerate}
    \item \textbf{Pre-trained Knowledge to Address Data Scarcity:}  
    A major challenge in blockchain data analysis is the scarcity of ground truth data or labeled datasets required for effective machine learning models. LLMs, trained on vast amounts of diverse data, bring extensive pre-trained knowledge that enables them to infer insights even in the absence of domain-specific datasets. This capability allows LLMs to act as robust tools for blockchain analytics, especially in scenarios where labeled data is limited or unavailable.

    \item \textbf{Generalizability Across Multiple Blockchains:}  
    Blockchains vary widely in their underlying protocols and data structures (e.g., UTXOs versus account models). Traditional analytical tools often require customization to be compatible with specific blockchain platforms. LLMs, however, exhibit high generalizability, enabling them to adapt to multiple blockchains without requiring extensive re-engineering. This makes LLMs particularly suited for environments where interoperability and scalability across heterogeneous blockchain networks are critical.

    \item \textbf{Explainability for Insightful Decision-Making:}  
    blockchain data analysis often generates complex insights that need to be interpreted by developers, auditors, and regulators. Explainability, one of the core strengths of LLMs, allows these models to provide understandable reasoning behind the outputs they generate. This capability not only builds trust in the insights derived from LLMs but also facilitates better decision-making by enabling users to understand the rationale behind key findings. In contexts such as fraud detection, compliance, and governance, explainability is essential for actionable insights and regulatory alignment.
\end{enumerate}

LLMs have shown promising results in specific areas of blockchain data analysis, such as smart contract auditing, where they help detect vulnerabilities and optimize code.
While such individual use cases exist, there is a lack of a unified framework to understand how LLMs can be effectively deployed across the diverse analytical tasks that blockchain ecosystems demand. 

This paper addresses this gap by presenting our positions on how LLMs-related techniques, such as prompt engineering and retrieval-augmented generation (RAG) and design patterns, can be leveraged in each downstream task of blockchain data analysis. Furthermore, key challenges such as cost, latency, and reliability remain underexplored, which could be prospective research challenges and opportunities in the future.

Our paper provides the following key contributions to this field.
\begin{enumerate} 
    \item \textbf{Comprehensive Framework for LLM Integration:} We present a systematic and comprehensive framework for integrating LLMs into blockchain data analysis, addressing diverse tasks such as fraud detection, smart contract auditing, market prediction, and governance evaluation.
    
    \item \textbf{Discussion of Techniques for Prompt Engineering:} We identify and discuss design patterns for prompt engineering, including in-context learning, RAG, reasoning frameworks (e.g., Chain-of-Thought (CoT)), and compression techniques for blockchain data analysis.
    
    \item \textbf{Discussion of Design Patterns:} We identify and discuss design patterns for LLM integration, categorizing them into roles such as enhancers and predictors and exploring their architectures, including workflows like data preprocessing, LLM-driven analysis, and hybrid approaches.

    \item \textbf{Broadening Use Case Perspectives:} We emphasize the versatility of LLMs in blockchain data analysis, from foundational tasks like anomaly detection to high-level applications such as decentralized finance (DeFi) monitoring and non-fungible token (NFT) analytics.

    \item \textbf{Identification of Research Directions:} We highlight six critical areas for future research in LLM-integrated blockchain data analysis, including latency, reliability, cost, scalability, generalizability, and autonomy, providing actionable insights for addressing these challenges.

\end{enumerate}
\begin{figure*}
    \centering
    \includegraphics[width=0.9\linewidth]{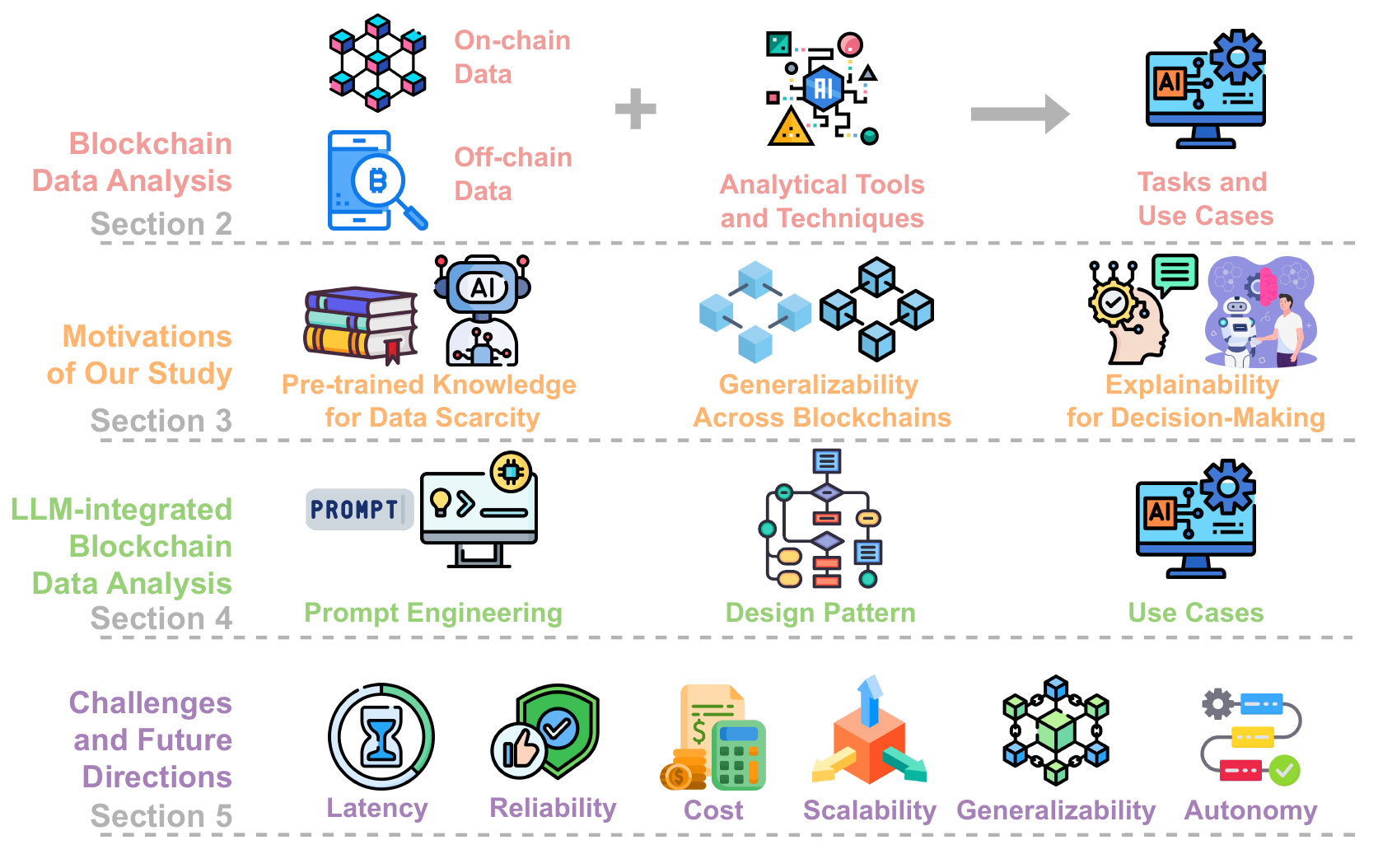}
    \caption{Paper Outline.}
    \label{fig:outline}
\end{figure*}
As illustrated in \figurename~\ref{fig:outline}, the remainder of the paper is organized as follows. 
Section~\ref{sec:blockchain_analysis} provides comprehensive data, downstream tasks on blockchain data analysis, underlying techniques, and its challenges.
Section~\ref{sec:motivations} states the motivations of our study.
Section~\ref{sec:LLM-integrated-blockchain-analysis} presents the proposed prompt mining techniques with RAG and systematic design patterns that make LLMs essential for blockchain data analysis.
Section~\ref{sec:challenges_and_future_directions} outlines the research opportunities and challenges associated with integrating LLMs into blockchain ecosystems. 
Finally, Section~\ref{sec:conclusion} concludes the paper.

\section{blockchain data analysis}
\label{sec:blockchain_analysis}

blockchain data analysis refers to the process of examining data recorded on blockchain networks and/or data available outside the blockchains to achieve downstream tasks, such as fraud detection, regulatory compliance, risk assessment in decentralized finance (DeFi), and smart contract auditing. Effective blockchain data analysis enables stakeholders, including developers, auditors, regulators, and financial institutions, to make data-driven decisions while safeguarding the network against malicious actors.

Due to its demand, a variety of blockchain analyses have been conducted. Though we will not fully cover the state-of-the-art as we already see many survey papers in this domain (e.g., \cite{Hou2021-survey-BDA, song2022survey-networks, Tovanich2021-survey-visualization, Qi2024-survey-graph}), here we summarize available data, tasks and use cases, and underlying techniques to achieve them.

\subsection{Data}
\begin{table*}[ht]
\centering
\caption{Comprehensive Data Types for Blockchain Analysis.}
\label{tab:comprehensive_data_types}
\resizebox{\linewidth}{!}{
\begin{tabular}{|p{1.5cm}|p{3cm}|p{3cm}|p{6cm}|p{4cm}|}
\hline
\textbf{Type} & \textbf{Sub-Type} & \textbf{Fields} & \textbf{Description} & \textbf{How to Capture} \\ \hline

On-Chain 
& Transaction Data 
    & Transaction ID 
    & Unique identifier for each transaction. 
    & Blockchain node RPC or third-party APIs (e.g., Etherscan, OpenSea) \\ \cline{3-4}
& & Sender and Receiver Addresses 
    & Wallet addresses involved in the transaction. 
    & \\ \cline{3-4}
& & Gas Fees 
    & Cost paid for executing the transaction. 
    & \\ \cline{2-4}
& Block Data 
    & Block Height 
    & Position of the block in the chain. 
    & \\ \cline{3-4}
& & Miner/Validator Information 
    & Details of the entity that mined or validated the block. 
    & \\ \cline{2-4}
& Smart Contract Bytecode 
    & Bytecode 
    & Compiled smart contract code stored on-chain. 
    & \\ \cline{2-4}
& Event Logs 
    & Event Name, Parameters, Block Number, Transaction Hash 
    & Logs emitted by smart contracts during execution, capturing activity such as token transfers or staking actions. 
    & \\ \cline{2-4}
& Token Data 
    & Token Transfers 
    & Details of token movements between addresses and their balances. 
    & \\ \cline{3-4}
& & NFT Metadata 
    & Metadata linked to non-fungible tokens (e.g., images, descriptions). 
    & \\ \hline

Off-Chain
& Smart Contract Source Code 
    & Solidity Code 
    & Human-readable source code defining smart contract logic. 
    & GitHub repositories or Etherscan \\ \cline{3-4}
& & ABI 
    & Interface used to interact with the contract. 
    & \\ \cline{2-5}
& Oracles and Market Data 
    & Price Data 
    & Historical and real-time prices of tokens. 
    & Third-party APIs (e.g., Binance, Coinbase, Chainlink) \\ \cline{3-4}
& & Trading Volume 
    & Volume of tokens traded over a given period. 
    & \\ \cline{2-5}
& Social Media and Community Data 
    & Forum Posts 
    & Discussions about blockchain projects and scams. 
    & Web scraping or APIs (e.g., Reddit, Bitcoin Forum, Tor) \\ \cline{3-5}
& & Microblogging Feeds 
    & Tweets mentioning blockchain topics. 
    & Twitter, Thread, Bluesky, Binance Square APIs\\ \cline{3-5}
& & Communication Channels 
    & Posts mentioning blockchain topics. 
    & Telegram, Discord APIs or bots \\ \cline{2-5}
& Regulatory and Legal Data 
    & Regulatory Guidelines 
    & Compliance standards such as AML and KYC. 
    & Government websites or compliance platforms \\ \cline{3-5}
& & Sanctions Lists 
    & Blacklisted addresses or entities. 
    & Compliance services (e.g., Chainalysis) \\ \cline{2-5}
& Developer Data 
    & Repository Activity 
    & Commits, pull requests, and issues in blockchain projects. 
    & GitHub APIs \\ \cline{3-5}
& & Roadmaps 
    & Development timelines and milestones. 
    & Project websites or GitHub \\ \hline

\end{tabular}
}
\end{table*}

Effective blockchain data analysis relies on a diverse range of data sources that provide critical insights into blockchain networks and applications. 
\tablename~\ref{tab:comprehensive_data_types} lists a comprehensive list of data for blockchain data analysis. It can be broadly categorized into \textit{on-chain data} and \textit{off-chain data}.

\subsubsection{On-chain Data}
On-chain data refers to information that can be fetched directly from blockchain networks. Key types of on-chain data include transactions, blocks, smart contract bytecodes, tokens, and network-side information. The on-chain data are typically captured via RPC (remote procedure call) to blockchain nodes or APIs to blockchain explorers or RPC providers, such as Infura, Alchemy, Quicknode, and Blockchain ETL.

\subsubsection{Off-chain Data}
Off-chain data refers to data that exists outside of the blockchain but is relevant for understanding and analyzing blockchain networks, such as smart contract source code, market data, social media conversations, and legal and regulation information. This data provides additional context and is often used to enrich on-chain data analysis. The off-chain data are mostly available via the APIs of social media, forums, and websites or scraping content from Tor (the onion router).

\subsection{Tasks and Use Cases}
This section summarizes a comprehensive list of key tasks that can be achieved by combining on-chain and off-chain data with underlying analytical techniques.

\subsubsection{Fraud Detection}
Fraud detection is a cornerstone of blockchain analytics, aimed at identifying and mitigating malicious activities on the blockchains \cite{Pranto2022-fraud-detection}. Fraudulent behaviors, such as phishing schemes (e.g., \cite{Chen2020-phishing, Chen2021-phishing}), Ponzi scams (e.g., \cite{Bartoletti2018-Ponzi, Toyoda2019-Ponzi}), and money laundering (e.g., \cite{Lorenz2020-laundering, Wang2024-laundering-graph}), exploit the pseudonymity and global nature of blockchain networks to evade detection. 
By leveraging techniques such as graph analysis, machine learning, and anomaly detection, fraud detection systems can uncover suspicious patterns in transaction flows, address interactions, and trading volumes. Additionally, integrating off-chain data, such as compliance reports and social media signals, provides a comprehensive view of fraudulent behaviors. 

\subsubsection{Smart Contract Analysis}
One of the most critical tasks in smart contract analysis is the detection of vulnerabilities that could compromise security and lead to significant financial or reputational losses. Depending on the use cases, the data to be analyzed are raw source code or compiled one (bytecode) and ABI. Techniques in NLP, graph mining, and software engineering are often used for smart contract analysis (e.g., \cite{Zhuang2021smart-contract, qian2022smart-contract, Chen2023chatgpt-smart-contract}). 

Fraudulent contracts, such as Ponzi schemes and ``honeypot'' scams, represent a significant risk in the blockchains (e.g., \cite{Chen2018-Ponzi-Ethereum}). These contracts are deliberately designed to deceive users, often by promising unsustainable returns or locking funds in ways that prevent withdrawals. 


Auditing smart contract logic ensures that the implemented functionality aligns with the project's stated goals and requirements \cite{xia2024auditgpt}. This process involves reviews of the contract's source code to verify that it behaves as intended under all possible conditions. For example, an auditing process may verify that a DeFi lending protocol correctly calculates interest rates, maintains collateralization thresholds, and prevents unauthorized fund withdrawals.




\subsubsection{Market Analysis and Prediction}
Market analysis and prediction are useful for understanding and forecasting trends in blockchain ecosystems, enabling investors and stakeholders to make informed decisions in volatile and dynamic markets. By leveraging historical on-chain data, such as transaction volumes and token flows, predictive models can anticipate price movements, identify trading opportunities, and assess market dynamics (e.g., \cite{jay2020price-prediction, khedr2021price-prediction}). Additionally, off-chain data sources, including social media sentiment, news articles, and public forums, are incorporated using NLP techniques to evaluate community sentiment and its impact on market behavior (e.g., \cite{abraham2018sentiment, parekh2022sentiment}). In DeFi, monitoring metrics such as total value locked (TVL), liquidity pool behavior, and staking activity, provides insights into the health and risks of financial protocols. Similarly, in the NFT space, tracking ownership trends, transaction activity, and market liquidity reveals user engagement and the evolving value of digital assets (e.g., \cite{ko2022NFT, nadini2021NFT}.

\subsubsection{Network, Governance, and Compliance Monitoring}
Network monitoring focuses on evaluating critical metrics such as node distribution, block propagation times, and transaction throughput to detect anomalies or vulnerabilities, such as network congestion or potential attacks (e.g., \cite{Zheng2018monitoring, Kim2021monitoring}). Governance monitoring assesses participation in decision-making processes, including on-chain voting and proposal outcomes, to ensure alignment with community objectives and detect potential centralization risks in decentralized governance systems \cite{Liu2021-governance}. Meanwhile, compliance monitoring involves tracking adherence to legal and regulatory frameworks, such as anti-money laundering (AML) and know-your-customer (KYC) requirements, by identifying suspicious transactions, sanction violations, or patterns indicative of money laundering. 

\subsubsection{Privacy Analysis}
Privacy analysis involves analyzing transactions conducted using privacy-focused cryptocurrencies such as Monero and Zcash (e.g., \cite{kumar2017Monero, moser2017Monero, kappos2018Zcash}). These coins employ advanced cryptographic techniques like ring signatures, stealth addresses, and zk-SNARKs to obfuscate transaction details, making them challenging to trace. Privacy analysis in this context focuses on identifying patterns in the use of such technologies, assessing their effectiveness, and ensuring compliance with regulations without compromising user anonymity. On the other hand, privacy analysis also involves detecting de-anonymization techniques or patterns that adversaries may use to infer identities or transactional relationships within blockchain networks. Techniques such as transaction graph analysis, address clustering, and metadata correlation can reveal vulnerabilities in seemingly private systems. 

\subsection{Analytical Tools and Techniques}
The following are the basic underlying techniques to achieve the above tasks with given data.

\subsubsection{Address Clustering}
Address clustering is a technical process in blockchain data analysis that aims to group blockchain addresses based on shared behavioral characteristics or transactional relationships, enabling the identification of entities and activity patterns. This process relies heavily on heuristic and algorithmic techniques to infer connections between addresses that are not explicitly linked due to the pseudonymous nature of blockchains (e.g., \cite{harrigan2016address-clustering, victor2020address-clustering, huang2023address-clustering}). For instance, many transactions on blockchains like Bitcoin produce a ``change'' address to return unspent outputs to the sender. Identifying such change addresses by analyzing output patterns and reusing addresses enables the clustering of multiple addresses likely controlled by the same entity.

\subsubsection{Supervised Machine Learning}
Supervised learning, mainly classification and regression, is used for prediction tasks where labeled data is available for training. Some examples include binary classification (e.g., fraudulent vs. non-fraudulent transactions), address classification into categories (e.g., exchange, Ponzi scheme, mixer), and price prediction. The underlying algorithms such as SVM (support vector machine) \cite{cortes1995SVM}, RF (Random Forests) \cite{breiman2001RF}, XGBoost \cite{chen2016xgboost}, and neural networks \cite{rumelhart1986NN}, are often used (e.g., \cite{Kilic2022-supervised-ML, ostapowicz2019-supervised-ML}).

\subsubsection{Unsupervised and Semi-supervised Learning}
Unsupervised and semi-supervised learning, such as time-series analysis, clustering, and anomaly detection, are employed when labeled data is unavailable, or only one class is available, helping to discover patterns or anomalies in blockchain data (e.g., \cite{Toyoda2018-time-series}).  
They could be statistical approaches like z-scores (e.g., \cite{montgomery2019z-score}) and proximity-based approaches (e.g., \cite{ide2009proximity}), and ML-based approaches like Isolation Forest \cite{liu2008isolation-forest} and One-Class (OC) SVM \cite{scholkopf2001-OC-SVM}.

On-chain data often includes time-stamped events, such as transactions and block creation times. Time-series analysis models temporal patterns to detect trends and anomalies. For instance, they could be used to identify change points in account behavior and model price movements or transaction volumes.

\subsubsection{Graph Mining and Network Analysis}
On-chain data can be represented as a graph, where nodes correspond to entities (e.g., wallet addresses) and edges represent relationships (e.g., transactions) \cite{akcora2022blockchain-networks}. Graph mining techniques are used to analyze the structure, patterns, and anomalies in these networks (e.g., \cite{khan2022graph}). When treating a blockchain network as a graph, we could apply techniques like community detection, centrality measurement, and shortest path analysis to detect high-activity addresses and the relationships between target entities (e.g., exchanges or coordinated fraud groups). We see trends in applying graph embedding techniques that convert graphs into low-dimensional representations for use in machine learning tasks (e.g., scammer address detection) \cite{Qi2024-survey-graph}.

\subsubsection{Natural Language Processing (NLP)}
Although primarily used for analyzing off-chain data, NLP plays a growing role in understanding smart contracts and blockchain-related discussions and documents. Sentiment analysis and Named Entity Recognition (NER) analyze public sentiment on social media or forums and extract mentions of projects, tokens, or addresses. 
Formal verification is a rigorous mathematical method used to prove the correctness and security of smart contracts or blockchain protocols by ensuring they meet predefined specifications and are free from vulnerabilities or logical errors.

For instance, these techniques are used for detecting Ponzi schemes operating on smart contracts (e.g., \cite{Chen2018-Ponzi-Ethereum}), correlating sentiment trends with market behavior (e.g., \cite{abraham2018sentiment, parekh2022sentiment}), identifying scams or phishing attempts in announcements \cite{Benetti2021-NLP}, and identifying vulnerabilities in smart contracts (e.g., \cite{bhargavan2016formal-verification, gao2021sverify}).

\subsubsection{Visualization}
Data visualization is a useful tool in blockchain data analysis, transforming complex and voluminous blockchain data into intuitive visual representations that facilitate understanding of blockchain networks \cite{Tovanich2021-survey-visualization}. Graph visualizations, for instance, effectively illustrate transaction flows and address clusters, revealing patterns such as fund movement between entities or coordinated behaviors in scams. Heatmaps highlight temporal trends, such as periods of heightened activity in token trading or network congestion during peak usage. Dashboards aggregate and display real-time metrics, including transaction throughput, gas fees, and validator performance, enabling stakeholders to monitor network health and activity (e.g., \cite{kuzuno2017visualization, yue2018visualization, Kinkeldey2019-dashboard}). Additionally, geospatial mapping can reveal the geographic distribution of blockchain nodes, providing insights into network decentralization and resilience.

\subsection{Challenges}
We want to emphasize that the existing blockchain data analysis papers more or less face the following challenges:

\begin{enumerate}
    \item \textbf{Pseudonymity:}
    Blockchain's pseudonymity, combined with obfuscation tools such as mixers and privacy coins, complicates the identification of fraudulent or illegal activities. Conventional methods often struggle to link suspicious addresses or transactions with real-world identities.

    \item \textbf{Lack of Labeled Datasets and Ground Truth:}
    The scarcity of labeled datasets limits the development of machine learning models for blockchain data analysis, making it difficult to detect new types of fraud or risks effectively.
    
    \item \textbf{Protocol Dependencies:}
    Blockchain networks operate independently with unique protocols and data structures (e.g., Bitcoin's UTXO model versus Ethereum's account-based model). This fragmentation makes it difficult to conduct cross-chain analyses and extract meaningful insights consistently.

    \item \textbf{Scalability Issues:}
    As the size of blockchain networks increases, the volume of transactional data grows exponentially. Analyzing large datasets in real time poses computational challenges, especially when multiple chains are involved.

    \item \textbf{Interpretability of Insights:}  
    Blockchain analytics tools often generate complex insights that require expertise to interpret. This creates a barrier for non-experts, such as regulators or general users, to engage with the network meaningfully.
\end{enumerate}

\section{Motivations of Our Study}
\label{sec:motivations}
\begin{figure}[t]
    \centering
    \includegraphics[width=\linewidth]{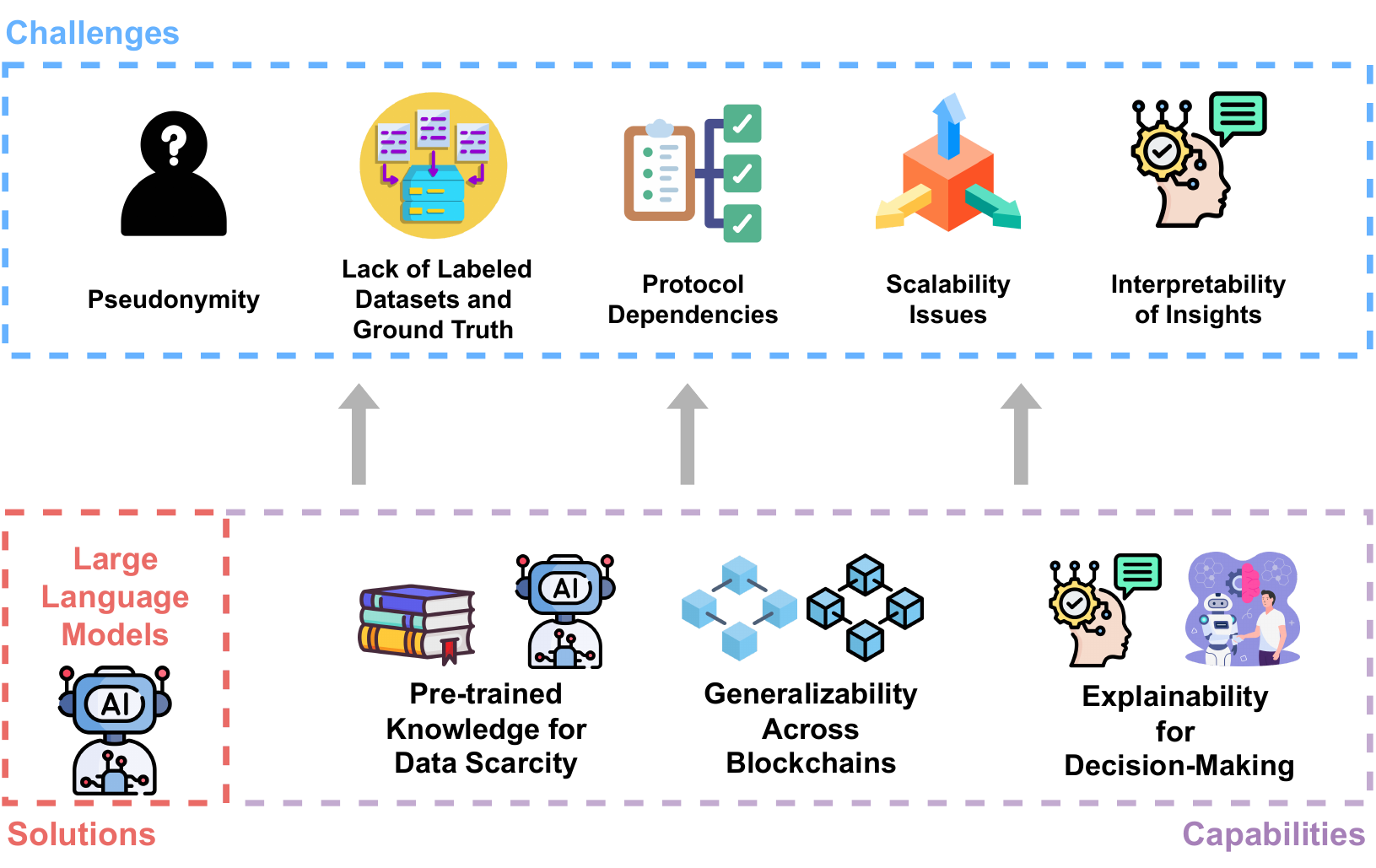}
    \caption{Illustration of How LLMs Could Solve Challenges in Blockchain Data Analysis.}
    \label{fig:llm4bcda}
\end{figure}

As illustrated in \figurename~\ref{fig:llm4bcda}, we foresee that LLMs will offer promising solutions to overcome some of these challenges through their advanced capabilities:

\begin{enumerate}
    \item \textbf{Pre-trained Knowledge for Data Scarcity:}  
    LLMs are trained on vast datasets across multiple domains, allowing them to generate meaningful insights even in the absence of blockchain-specific labeled datasets. This capability addresses the challenge of limited ground truth by leveraging knowledge from related contexts.

    \item \textbf{Generalizability Across Blockchains:}  
    LLMs can understand and process information from diverse blockchain protocols, enabling cross-chain analysis without the need for extensive re-engineering. Their adaptability makes them ideal for environments where multiple blockchains coexist with different architectures and consensus mechanisms.

    \item \textbf{Explainability for Decision-Making:}  
    LLMs excel at providing human-readable explanations for complex insights, facilitating better decision-making. This feature is critical for building trust in blockchain systems, as it allows auditors, regulators, and developers to understand the reasoning behind detected patterns or recommendations.

\end{enumerate}

We argue that a thorough discussion of available data and design patterns for LLM-integrated analysis, along with an exploration of how such combinations achieve downstream tasks, is currentlywmissing. While there are related survey papers on LLM integration in fields such as time-series analysis \cite{Jin2024-LLM-time-series-survey} and graph analysis \cite{Qi2024-survey-graph, Fan2024-survey-graph, Chen2023-survey-graph}, to the best of our knowledge, a comprehensive and systematic paper on the integration of LLMs in blockchain data analysis has yet to be published.

This paper seeks to fill this gap by providing a comprehensive framework for understanding and leveraging how LLMs could benefit downstream tasks in blockchain data analysis\footnote{Though it is an interesting topic, we will not cover how blockchain can benefit AI and LLMs. Please refer to some survey papers (e.g., \cite{luo2024bc4llm, geren2024BC-for-LLM, witt2024BC-AI}) for this topic.}. By systematically exploring the diverse types of data available for blockchain data analysis and the architectural approaches for integrating LLMs, we aim to highlight how these powerful models can enhance analytical capabilities, improve data-driven decision-making, and address key challenges in areas such as fraud detection, smart contract security, compliance monitoring, and market prediction.

\section{LLM-integrated blockchain data analysis}
\label{sec:LLM-integrated-blockchain-analysis}

LLMs are advanced machine learning models designed to understand, generate, and manipulate natural language. They are typically transformer-based pre-trained models containing billions of parameters \cite{vaswani2017attention}. Trained on massive datasets across diverse domains, LLMs such as GPT-4 have become powerful tools for various applications, including text generation, summarization, and reasoning. 

The performance of LLMs is heavily influenced by the way input prompts are structured. Techniques such as prompt engineering, in-context learning, RAG, and reasoning frameworks help guide the model in producing accurate and relevant outputs. Also, design patterns, i.e., how LLMs should be integrated into blockchain data analysis, have not yet been discussed. This section covers major techniques in prompt engineering and our suggested systematic classification of the design patterns in LLMs-integrated blockchain data analysis.

\subsection{Prompt Engineering}
Prompt engineering is a critical aspect of LLM applications, as it directly influences the model's ability to understand tasks, generate accurate outputs, and adapt to diverse use cases by carefully crafting inputs that guide its reasoning and behavior.

\subsubsection{Basics of Prompting}
We start with the basics of prompting. First, we need to identify the objectives with analytical tasks and desired outcomes, for example, address classification and price prediction with reasoning and fraud reporting generation. 
We can then design a template that provides clear and concise instructions for the LLM based on the defined objectives. 
\begin{itemize}
    \item \textbf{Instruction:} A clear statement of the task or question.
    \item \textbf{Context:} Supporting information or retrieved data (e.g., past knowledge about fraud).
    \item \textbf{Input Data:} The raw data to be analyzed (e.g., the list of transactions).
    \item \textbf{Formatting Guide:} Optional guidelines specifying the expected format of the output (e.g., numerical scores, textual explanations, or predefined categories).
\end{itemize}

The following shows the basic structure of a prompt template. 
For ease of understanding, we will take an example of a simple fraud detection use case from now on. The template should include placeholders for dynamic data and specify the context of the task in code (e.g., in Python). For instance:
\begin{tcolorbox}[prompt]
\begin{prompt}
Analyze the following transaction history and identify risk level from 0 to 1 scale (0: no risk, 1: highest risk): 
Transaction Data: \Code{\{transaction\_data\}}
Risk level:
\end{prompt}
\end{tcolorbox}

Transaction history captured from a blockchain, e.g., a list of JSON-formatted transactions, is embedded in \texttt{\Code{\{transaction\_data\}}} to complete a prompt.
We could manually craft templates or use an LLM to generate ones. This process often requires iteratively refining and experimenting with prompts to optimize the quality of the generated responses. 

\subsubsection{In-Context Learning}
We could extend the above template to accommodate more contextual information. In-context learning, or few-shot learning, involves providing an LLM with a carefully designed prompt containing examples (demonstrations) of the task at hand, followed by a query. The model uses the examples in the context to infer the desired task and generate the correct response. Here is a template example with two demonstrations. An LLM analyzes the given transaction data in the query and outputs its decision and explanation based on the two demonstrations above.
\begin{tcolorbox}[prompt]
\begin{prompt}
### Instruction ###
You are an expert in blockchain fraud detection. Analyze the given transaction data and determine whether it is likely fraudulent. Provide an explanation for your decision.

\color{teal}{### Examples ###}

Transaction Data:
\{
  "Transaction ID": "tx12345",
  "Sender Address": "0xA1B2C3D4E5",
  "Receiver Address": "0x123ABC456DEF",
  "Value": "10 ETH",
  "Timestamp": "2024-11-25T14:30:00Z",
  "Notes": "Sent to a known exchange address."
\}
Decision: Non-Fraudulent
Explanation: The receiver address is a known exch-
ange wallet, and the value is within typical tran-
saction limits.

---

Transaction Data:
\{
  "Transaction ID": "tx67890",
  "Sender Address": "0x987654321ABC",
  "Receiver Address": "0x654321FEDCBA",
  "Value": "500 ETH",
  "Timestamp": "2024-11-26T08:15:00Z",
  "Notes": "Unusual activity flagged: large trans-
  action to an unknown wallet."
\}
Decision: Fraudulent
Explanation: The transaction value is significantly
large, and the receiver address is unknown, which raises suspicion of potential fraud.

---
\end{prompt}
\end{tcolorbox}

\begin{tcolorbox}[prompt]
\begin{prompt}
### Query ###

Transaction Data:
\{
  "Transaction ID": "tx54321",
  "Sender Address": "0xABCDE12345F",
  "Receiver Address": "0xFEDCBA67890",
  "Value": "0.5 ETH",
  "Timestamp": "2024-11-26T10:45:00Z",
  "Notes": "Transaction from a personal wallet to another wallet."
\}
\Code{Decision:}
\Code{Explanation:}
\end{prompt}
\end{tcolorbox}

\subsubsection{Retrieval-Augmented Generation (RAG)}
In the in-context learning example, we assumed we already had contextual information related to the question. However, such insights are often stored in a large knowledge database, and it is infeasible to embed every single insight in the database.
The basic idea of RAG is to retrieve the most relevant contexts from such a database using similarity search and append them in a prompt. To illustrate, consider a fraud detection task where external metadata is retrieved and incorporated:
    
\begin{tcolorbox}[prompt]
\begin{prompt}
You are an expert in blockchain analysis. Using the provided external metadata and transaction history, identify whether the transaction is likely fraudulent.

External Metadata: \Code{\{retrieved\_data\}}
Transaction History: \Code{\{transaction\_data\}}

Answer with "Fraudulent" or "Non-Fraudulent" and provide reasoning.
\end{prompt}
\end{tcolorbox}

Assuming we have domain knowledge about scammers (e.g., scammers' transaction patterns), we search for such information in a database and append it in \texttt{\Code{\{retrieved\_data\}}}.

\subsubsection{Reasoning Frameworks}
Some analytical use cases require intricate reasoning. By reasoning step-by-step, the LLM was found to reduce errors and handle complexity more effectively (e.g., \cite{kojima2022step-by-step, Ganesh2024-patterns-LLM}).
Reasoning frameworks, such as Chain-of-Thought (CoT), Tree-of-Thought (ToT), and Graph-of-Thought (GoT), guide LLMs to generate intermediate steps or explore decision paths rather than directly jumping to an answer.

CoT involves a linear progression of thoughts (sequential reasoning), while ToT organizes reasoning in a tree-like structure to explore multiple options at each step (hierarchical reasoning), and GoT leverages graph-like relationships to handle dependencies or multiple paths simultaneously (interconnected reasoning).

Due to space limitations, we put the examples of CoT, ToT, and GoT in the fraud detection scenarios in Appendix~\ref{sec:examples_of_reasoning_frameworks}.

\subsubsection{Compress Embedded Data}
Data compression techniques aim to reduce the token footprint of large inputs, such as transaction histories or long smart contract logs, enabling efficient processing within LLM token limits. The following are some techniques for compression.

\begin{itemize}
    \item \textbf{Feature Extraction and Summarization:} Extract distinguished features from transactions, such as frequency of transactions, transaction volume per address, and time intervals, and provide them in a prompt rather than raw transactions.
    \item \textbf{Sampling:} For historical data, retain only the most recent or relevant portions of the input that align with the tasks.
    \item \textbf{Format Conversion:} Reduce the context size by converting the format (e.g., from JSON to plain text).
\end{itemize}

\subsection{Design Patterns}
\begin{figure*}[ht!]
    \centering
    \resizebox{\linewidth}{!}{ 
        \begin{minipage}{\linewidth}
            \centering
            \begin{subfigure}[b]{0.49\linewidth}
                \centering
                \includegraphics[width=\linewidth]{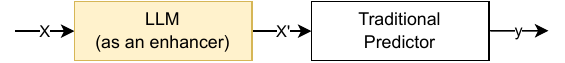}
                \caption{Pattern 1}
                \label{fig:design-pattern-1}
            \end{subfigure}
            \begin{subfigure}[b]{0.49\linewidth}
                \centering
                \includegraphics[width=\linewidth]{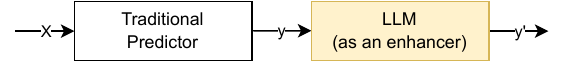}
                \caption{Pattern 2}
                \label{fig:design-pattern-2}
            \end{subfigure}
            
            \vskip\baselineskip
            \begin{subfigure}[b]{0.49\linewidth}
                \centering
                \includegraphics[width=.55\linewidth]{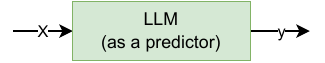}
                \caption{Pattern 3}
                \label{fig:design-pattern-3}
            \end{subfigure}
            \begin{subfigure}[b]{0.49\linewidth}
                \centering
                \includegraphics[width=\linewidth]{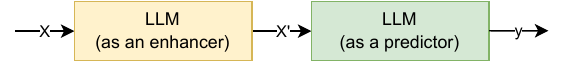}
                \caption{Pattern 4}
                \label{fig:design-pattern-4}
            \end{subfigure}
            
        \end{minipage}
    }
    \caption{The Proposed Design Patterns that Incorporate LLMs into Blockchain Analysis.}
    \label{fig:design-patterns}
\end{figure*}

Integrating LLMs into blockchain data analysis introduces solid design patterns that cater to diverse analytical needs. Following the classification in \cite{Jin2024-LLM-time-series-survey}, these patterns can be broadly classified into two categories: \textit{LLM-as-enhancers}, where LLMs augment specific tasks while the final prediction or decision is made by other components, and \textit{LLM-as-predictors}, where LLMs directly output the final prediction. 
As shown in \figurename~\ref{fig:design-patterns}, these patterns can be further categorized into four approaches based on how LLMs interact with input data ($X$), intermediate representations ($X'$), and final outputs ($y$). This section describes these patterns, provides their rationales, and highlights their applicability in blockchain data analysis.

Whichever the pattern, we should follow the data collection and the prompt-engineering steps above to control the input $X$ or $X'$ and the output $X'$ or $y$.

\subsubsection*{Pattern 1: $X \rightarrow$ [LLM] $\rightarrow X' \rightarrow$ [Traditional Predictor] $\rightarrow y$}

In this pattern, LLMs are employed as \textit{enhancers} to preprocess or augment the input data ($X$) into an enriched representation ($X'$), which is subsequently analyzed using traditional techniques to generate the final output ($y$). 
This approach leverages the LLM's ability to extract contextual information or transform raw data into structured features, enabling traditional methods to perform downstream tasks more effectively.

\noindent \textbf{Examples}:

Several research works demonstrate the effective application of Pattern 1. 
Yu et al. employ a BERT-based LLM to process blockchain transaction data ($X$) into enriched token representations ($X'$), which are then analyzed using a reconstruction error-based detector for anomaly classification ($y$) \cite{Yu2024BlockFound}. Another example we found is Rahman et al.'s work that utilizes pre-trained language models (DistilBERT, MiniLM, and FLAN-T5) to transform cryptocurrency-related social media text ($X$) into contextual embeddings ($X'$), followed by traditional classification methods to determine sentiment polarity ($y$) \cite{wahidur2024zero-shot-sentiment}. Both studies exemplify how LLMs can serve as sophisticated feature extractors, enriching the input data with contextual information before applying conventional prediction methods.

\subsubsection*{Pattern 2: $X \rightarrow$ [Traditional Predictor] $\rightarrow y \rightarrow$ [LLM] $\rightarrow y'$}

This pattern applies traditional analysis techniques to process the input data ($X$) and generate an intermediate output ($y$), which is then refined or interpreted by the LLM to produce the final output ($y'$).
Traditional methods handle feature extraction or aggregation efficiently, while LLMs excel at complex reasoning, summarization, or generating human-readable outputs.

\noindent \textbf{Examples}:

LLM-as-predictors are systems where LLMs directly generate the final prediction or decision, often utilizing their pre-trained knowledge and reasoning capabilities. PropertyGPT leverages Slither, a static analysis tool, to first extract dependency data from smart contracts ($X \rightarrow$ [Slither] $\rightarrow y$), then employs GPT-4o to refine and verify these dependencies through multiple roles including checker, evaluator, and verifier ($y \rightarrow$ [GPT-4o] $\rightarrow y'$) \cite{Liu2024PropertyGPT}. This approach achieved significant improvements in precision (0.91 vs 0.88) and recall (0.96 vs 0.84) compared to using the Slither tool alone. Similarly, Ren and Wei proposed Sligpt that utilizes Slither as an initial analyzer to collect state variables and dependency information ($X \rightarrow$ [Slither] $\rightarrow y$), followed by GPT-4o's reasoning capabilities to refine the analysis results through a chain-of-thought process ($y \rightarrow$ [GPT-4o] $\rightarrow y'$) \cite{RenWei2024Sligpt}. The evaluation showed this combined approach outperformed both standalone tools, achieving an F1 score of 0.93 compared to 0.86 for Slither and 0.80 for GPT-4o alone.

\subsubsection*{Pattern 3: $X \rightarrow$ [LLM] $\rightarrow y$}

In this pattern, the LLM directly processes the input data ($X$) to produce the final output ($y$), functioning as the primary analytical component. This approach leverages the LLM's pre-trained knowledge and reasoning capabilities, enabling it to perform both feature extraction and prediction without auxiliary methods.

\noindent \textbf{Examples}:

Pattern 3 is exemplified by several recent works that employ LLMs as end-to-end solutions for blockchain security and fraud detection. Gai et al. implemented a dynamic, real-time approach called BlockGPT by directly processing blockchain transaction traces through a pre-trained LLM to detect anomalous activities \cite{Gai2023BlockGPT}. The model ingests raw transaction data and produces anomaly scores without requiring intermediate feature engineering or auxiliary models. Similarly, Hu et al. demonstrated this pattern by utilizing a pre-trained transformer to directly analyze transaction sequences for fraud detection \cite{Hu2023BERT4ETH}. The model processes transaction features and generates fraud predictions through direct end-to-end learning. These works showcase how LLMs can serve as comprehensive analytical engines that directly transform raw blockchain data into security insights, exemplifying the $X \rightarrow$ [LLM] $\rightarrow y$ pattern where the LLM serves as the primary computational component without requiring additional processing steps or auxiliary models.

\subsubsection*{Pattern 4: $X \rightarrow$ [LLM] $\rightarrow X' \rightarrow$ [LLM] $\rightarrow y$}

This pattern employs a multi-stage architecture where an initial LLM processes input data ($X$) into an intermediate representation ($X'$), which is further refined or analyzed by a second LLM to produce the final output ($y$). Multi-stage LLM designs allow specialization in distinct tasks, such as initial summarization followed by in-depth reasoning, enabling complex workflows requiring iterative refinement.

\noindent \textbf{Examples}:
Recent research exemplifies Pattern 4's multi-stage LLM architecture through several notable works. Luo et al. implemented a two-stage process where the first LLM analyzes smart contract code to construct call graphs and Contract-External Function-Call (CEC) files, followed by a second LLM that leverages these intermediate representations to detect specific vulnerabilities \cite{Luo2024FELLMVP}. This demonstrates how the intermediate representation $X'$, annotated CEC files, enhances the final vulnerability detection accuracy. Similarly, Mothukuri et al. employed a chain where one LLM generates the Control Flow Graph (CFG) annotations, while a subsequent LLM called LLMGraphAgent utilizes these annotations to identify security vulnerabilities and suggest fixes \cite{Mothukuri2024LLMSmartSec}. Li et al. proposed a method called CryptoTrade that first uses an LLM to process market and news data into analytical reports ($X'$), which are then fed into a second LLM that generates specific trading decisions \cite{li2024CryptoTrade}. These implementations demonstrate how Pattern 4's staged approach enables specialized processing at each step, with the intermediate representation ($X'$) serving as a refined input that enhances the final output's accuracy.



\subsection{Use Cases}
Here, we discuss how they would improve the operations from a use-case point of view. We try to be aligned with the use cases in Section~\ref{sec:blockchain_analysis}.

\subsubsection{Fraud Detection}
LLMs will demonstrate significant potential in enhancing fraud detection capabilities on blockchain networks and applications. 
One prominent application is the development of sophisticated account representation models. Hu et al.'s BERT4ETH \cite{Hu2023BERT4ETH} is a pre-trained transformer model specifically designed for Ethereum fraud detection tasks. This model utilizes masked address prediction as a pre-training task to capture the co-occurrence relationships between transactions, enabling it to generate more expressive and context-aware account representations compared to traditional graph-based methods. BERT4ETH demonstrates superior performance in detecting phishing accounts and linked accounts that are being de-anonymized. 
For instance, in phishing account detection, BERT4ETH achieved an F1 score improvement of 21.61 absolute percentage points over the best-performing graph neural network model. This significant enhancement in detection accuracy showcases the potential of LLMs in addressing the challenges posed by sophisticated fraud schemes on blockchain platforms.

Another application is the ZipZap framework proposed by Hu et al.~\cite{ZipZap}. ZipZap addresses the computational challenges associated with training LLMs in large-scale blockchain datasets. By incorporating frequency-aware compression techniques and an asymmetric training paradigm, ZipZap achieves both parameter efficiency and computational efficiency in LM training for blockchain applications. ZipZap's frequency-aware compression technique allows for a remarkable 92.5\% reduction in model parameters with only a marginal performance loss. This compression is achieved by correlating the embedding dimension of an address with its occurrence frequency in the dataset, effectively addressing the power-law distribution of address frequencies in blockchain transactions. Such efficiency improvements are crucial for the deployment of LLM-based fraud detection systems on a scale, particularly given the ever-growing volume of blockchain data.

The application of LLMs in blockchain fraud detection offers several key advantages. LLMs can leverage their pre-trained knowledge to generate meaningful insights even when blockchain-specific labeled datasets are limited. This capability is particularly valuable in the rapidly evolving blockchain ecosystem, where new fraud patterns may emerge faster than labeled data can be collected. Although the discussed models focus on Ethereum, the underlying principles of LLM-based fraud detection can be adapted to other blockchain platforms. This generalizability allows for cross-chain analysis without extensive re-engineering, making LLMs a versatile tool for fraud detection across diverse blockchain ecosystems. 

\subsubsection{Smart Contract Analysis}
Smart contract analysis is one of the intuitive use cases that LLMs can contribute to. This domain is rather advanced in terms of LLM integration \cite{He2024-LLM-BC-security-survey}.
LLMs have been extensively studied to identify and repair vulnerabilities within smart contracts. Liu et al. introduced FELLMVP, an ensemble LLM framework to classify vulnerabilities in smart contracts~\cite{Luo2024FELLMVP}. This approach combines multiple LLM agents, each specialized in distinct areas of security auditing, including contract code analysis, vulnerability identification, and security summary. By leveraging the collective capabilities of these specialized agents, FELLMVP demonstrates superior performance in detecting a wide range of vulnerabilities, including complex logic vulnerabilities that traditional tools often overlook. Sun et al. developed two innovative tools: ACFIX and GPTScan \cite{Wei2024LLMSmartAudit}. ACFIX utilizes GPT-4 to repair access control vulnerabilities in smart contracts. By mining common role-based access control (RBAC) practices and guiding the LLM with contextual information, ACFIX achieves a high success rate in repairing vulnerabilities, significantly outperforming baseline models. GPTScan integrates GPT with static analysis to detect logic vulnerabilities in smart contracts. This approach showcases the ability of LLMs to address the challenge of data scarcity by utilizing pre-trained knowledge to generate meaningful insights even with limited blockchain-specific labeled datasets.

LLMs have also been employed to improve formal verification processes for smart contracts. Liu et al. proposed PropertyGPT, a novel system that leverages LLMs for the automated generation of smart contract properties \cite{Liu2024PropertyGPT}. This method enables the generation of diverse types of properties, including invariants, pre-/post-conditions, and rules, significantly enhancing the effectiveness of formal verification. PropertyGPT addresses several challenges in property generation, ensuring that the generated properties are compilable, appropriate, and runtime-verifiable. It uses compilation and static analysis feedback as an external oracle to guide LLMs in iteratively revising the generated properties. This approach not only improves the quality of generated properties but also demonstrates the potential of LLMs to transfer knowledge from existing human-written properties to new, unknown smart contract codes.

Sligpt is a methodology that integrates GPT-4o with the static analysis tool Slither to perform data dependency analyses on Solidity smart contracts \cite{RenWei2024Sligpt}. This approach not only improves the accuracy of code analysis but also enables users to query and analyze smart contracts using natural language, significantly enhancing the user experience and accessibility of smart contract analysis tools.

\subsubsection{Market Analysis and Prediction}
LLMs have shown significant potential to analyze the sentiment of social networks and news sources to predict cryptocurrency price movements. Using their natural language understanding capabilities, these models could process large amounts of textual data to gauge market sentiment and its potential impact on cryptocurrency values. Roumeliotis et al. utilized GPT-4, BERT, and FinBERT models to perform sentiment analysis on cryptocurrency news articles for predicting Ethereum price trends \cite{Roumeliotis2024sentiment-analysis}. This approach not only captures the sentiment of the news but also identifies correlations between different cryptocurrencies, providing valuable information for investment decisions. By elucidating the reasoning behind sentiment assessments and price predictions, these models can enhance transparency and trust in cryptocurrency markets, aiding investors and regulators in decision-making processes.

LLMs can analyze complex relationships and correlations between multiple cryptocurrencies. Singh and Bhat developed a Transformer-based neural network model to predict Ethereum prices \cite{Singh2024-sentiment-analysis}. The model incorporated not only Ethereum's own price and volume data but also data from other highly correlated cryptocurrencies such as Polkadot and Cardano. This approach leverages the interdependencies between different assets in the cryptocurrency market, aiming to improve prediction accuracy. This approach showcases the generalizability of LLMs across different blockchain ecosystems, allowing for comprehensive market analysis without the need for extensive reengineering for each cryptocurrency. The ability to process and correlate data from multiple sources demonstrates the potential of LLMs to provide a holistic view of the cryptocurrency market.

LLMs can enhance the analysis of on-chain transaction data. Li et al.'s LLM-driven trading agent, CryptoTrade, integrates both on-chain and off-chain data to optimize cryptocurrency trading decisions \cite{li2024CryptoTrade}. By analyzing transaction statistics, market data, and news summaries, the model can make more informed trading decisions. This application highlights the potential of LLMs to address data scarcity issues in blockchain analytics. By leveraging pre-trained knowledge from diverse domains, LLMs can generate meaningful insights even when blockchain-specific labeled datasets are limited. 

\subsubsection{Network, Governance, and Compliance Monitoring}
LLMs will be a powerful tool for anomaly detection in blockchain networks, offering several advantages over traditional methods. 
One notable application in this domain is BlockFound, a customized foundation model for anomaly detection in blockchain transactions \cite{Yu2024BlockFound}. BlockFound models the unique multi-modal data structure of blockchain transactions, which typically contain blockchain-specific tokens, texts, and numbers. The model employs a modularized tokenizer to handle these diverse inputs, effectively balancing information across different modalities. BlockFound's approach addresses key challenges in blockchain anomaly detection by utilizing pre-trained knowledge to generate meaningful insights even with limited blockchain-specific labeled datasets. This capability is crucial in the rapidly evolving blockchain landscape, where new types of anomalies may emerge faster than labeled data can be collected. The model also demonstrates effectiveness on both Ethereum and Solana networks, showcasing its ability to adapt to different blockchain architectures without extensive reengineering. This cross-chain applicability is particularly valuable in the increasingly interconnected blockchain ecosystem. 

Though not using a proprietary LLM, another notable approach in this field is BlockGPT \cite{Gai2023BlockGPT}, which uses a GPT-style model to detect anomalous blockchain transactions. BlockGPT generates tracing representations of blockchain activity and trains an LLM from scratch to act as a real-time blockchain anomaly detection. This method offers an unrestricted search space and does not rely on predefined rules or patterns, enabling it to detect a wider range of anomalies.

These LLM-based methods offer several advantages over traditional anomaly detection techniques. They can capture complex patterns and contextual information in transaction data, allowing for more nuanced anomaly detection compared to rule-based systems. By learning from vast amounts of data, LLMs can potentially identify novel types of anomalies that might be missed by static, rule-based systems. 

\subsubsection{Privacy Analysis}
To the best of our knowledge, we could not find any papers that use LLMs for transaction privacy analysis. However, we believe it could still benefit from LLMs. By integrating on-chain and off-chain data, LLMs may help identify anomalous behaviors in privacy coins like Monero or Zcash. Furthermore, LLMs' inherent capabilities of finding patterns could be useful for detecting de-anonymization techniques, such as new heuristics of address clustering.

\section{Challenges and Future Research Directions}
\label{sec:challenges_and_future_directions}

To fully realize the potential of Large Language Models (LLMs) in blockchain data analysis, we argue that future research must address six critical areas, namely (1) latency, (2) reliability, (3) cost, (4) scalability, (5) generalizability, and (6) autonomy. 
We believe that research in these areas will pave the way for a new generation of LLM-powered tools, transforming the landscape of blockchain analytics with greater efficiency, reliability, and innovation.

\subsection{Latency}
First, latency remains a key challenge, as the responsiveness of LLMs must be optimized for real-time blockchain applications such as fraud detection, compliance monitoring, and DeFi trading. 
Hence, it is crucial to design latency-aware methods. For instance, one potential idea is a hybrid LLM design that offloads time-consuming tasks, such as complex reasoning, keeps the insights in a local database, and uses them with a local LLM to obtain the final results. Here it is important to carefully design systems as we presented above.

\subsection{Reliability}
Second, improving reliability is essential, particularly in mitigating hallucinations and ensuring that LLM outputs are accurate, consistent, and aligned with domain-specific knowledge. 
Alignment techniques \cite{Ji2023-alignment-survey}, such as reinforcement learning from human feedback (RLHF), critique models (e.g., \cite{gao2024smart-contract-bugs}), could improve reliability.

\subsection{Cost}
Third, reducing the cost of deploying and maintaining LLMs is crucial. When LLMs are locally executed, GPU resources would be costly. On the other hand, when LLMs are executed remotely via service providers' endpoints (e.g., OpenAI, HuggingFace), API charges would be costly. It is typically advisable to start with ``small'' language models and increase the model size until finding a sweet spot that balances output quality and cost. We believe that this is case-by-case and tuned accordingly based on the problem at hand.

\subsection{Scalability}
Fourth, scalability needs further exploration, including techniques like data compression in prompt engineering and modular frameworks to handle the ever-increasing volume of blockchain data, which could also reduce costs and latency. 

\subsection{Generalizability}
Fifth, enhancing generalizability is vital to ensure that LLMs can adapt across diverse blockchain platforms, protocols, and use cases, leveraging pattern extraction and foundation model principles for robust cross-chain analysis. For instance, prompt engineering focusing on blockchain-agnostic transaction pattern analysis could be useful.

\subsection{Autonomy}
Finally, achieving autonomy through the development of AI agents that can independently retrieve, analyze, and act on blockchain data will enable fully automated workflows, minimizing the need for human intervention in complex decision-making processes. 

\begin{figure}[t]
    \centering
    \includegraphics[width=\linewidth]{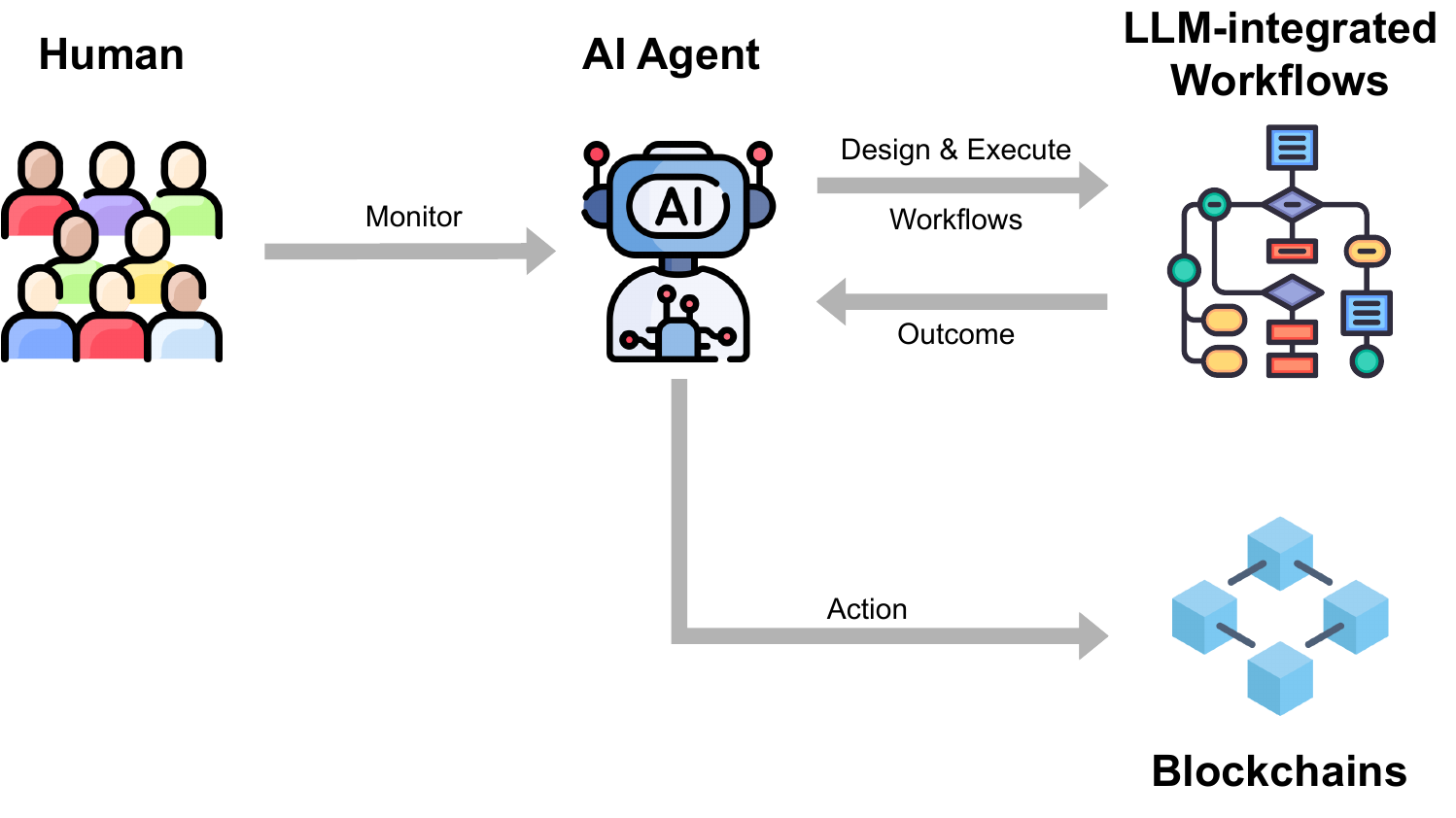}
    \caption{An AI Agent that Automates Decision Making on Blockchains.}
    \label{fig:ai-agent}
\end{figure}
It may sound ambitious, but it would be useful if AI agents could build a workflow toolchain based on the users' inputs as in \figurename~\ref{fig:ai-agent}, automating tasks such as network monitoring, fraud detection, and trading. To achieve this, we believe most of the above areas have to be enhanced.

\section{Conclusion}
\label{sec:conclusion}
This paper has outlined the potential of LLMs in blockchain data analysis. We have argued that by leveraging their pre-trained knowledge, generalizability, and explainability, LLMs can augment traditional analysis techniques, enabling more robust and scalable solutions. We have presented comprehensive available data, prompt design, and system design patterns to enhance the capabilities in various use cases.

Despite their promise, integrating LLMs into blockchain workflows presents challenges such as latency, reliability, cost, scalability, generalizability, and autonomy. Addressing these areas requires concerted research efforts, including optimizing prompt engineering, developing anti-hallucination methods, and building AI agents capable of autonomous decision-making. 

We believe that by overcoming existing limitations and fostering interdisciplinary collaboration, integrating LLMs into blockchain analytics can drive the next wave of innovation, transforming how blockchains are understood, secured, and utilized.

\bibliographystyle{ieeetr}
\bibliography{ref}

\appendices
\section{Examples of Reasoning Frameworks}
\label{sec:examples_of_reasoning_frameworks}
\subsection{Chain-of-Thought (CoT)}
CoT prompting involves encouraging an LLM to break down its reasoning process into intermediate steps.
\begin{tcolorbox}[prompt]
\begin{prompt}
### Instruction ###
You are an expert in blockchain fraud detection. For each transaction, analyze the data step-by-step to determine if the transaction is fraudulent or not. Provide a clear explanation for your decision.

### Example 1 ###

Transaction Data:
{
  "Transaction ID": "tx12345",
  "Sender Address": "0xA1B2C3D4E5",
  "Receiver Address": "0x123ABC456DEF",
  "Value": "10 ETH",
  "Timestamp": "2024-11-25T14:30:00Z",
  "Notes": "Sent to a known exchange address."
}

Reasoning:
1. The transaction involves a sender transferring 10 ETH to a receiver.
2. The receiver address matches a known exchange wallet.
3. The transaction value is within typical limits for exchange deposits.
4. No unusual activity or flags are associated with the transaction.

Decision: Non-Fraudulent
Explanation: The receiver is a known exchange address, and the transaction value is reasonable.

---

### Example 2 ###

Transaction Data:
{
  "Transaction ID": "tx67890",
  "Sender Address": "0x987654321ABC",
  "Receiver Address": "0x654321FEDCBA",
  "Value": "500 ETH",
  "Timestamp": "2024-11-26T08:15:00Z",
  "Notes": "Unusual activity flagged: large transaction to an unknown wallet."
}

Reasoning:
1. The transaction involves a sender transferring 500 ETH to a receiver.
2. The receiver address is unknown and not associated with a verified entity.
3. The transaction value is exceptionally large compared to average transactions.
4. The notes indicate unusual activity, further raising suspicion.

Decision: Fraudulent
Explanation: The transaction is flagged due to the high value and the unknown receiver, which suggests potential fraud.

---

### Query ###

Transaction Data:
{
  "Transaction ID": "tx54321",
  "Sender Address": "0xABCDE12345F",
  "Receiver Address": "0xFEDCBA67890",
  "Value": "0.5 ETH",
  "Timestamp": "2024-11-26T10:45:00Z",
  "Notes": "Transaction from a personal wallet to another wallet."
}

Reasoning:

\end{prompt}
\end{tcolorbox}

\subsection{Tree-of-Thought (ToT)}
ToT prompting involves guiding the model to generate a hierarchical reasoning process where multiple branches of reasoning are explored at each step. 
\begin{tcolorbox}[prompt]
\begin{prompt}
### Instruction ###
You are an expert in blockchain fraud detection. For each transaction, explore multiple possibilities at each step of the reasoning process to determine if the transaction is fraudulent. Evaluate and refine the paths before arriving at a decision.

### Example 1 ###

Transaction Data:
{
  "Transaction ID": "tx67890",
  "Sender Address": "0x987654321ABC",
  "Receiver Address": "0x654321FEDCBA",
  "Value": "500 ETH",
  "Timestamp": "2024-11-26T08:15:00Z",
  "Notes": "Unusual activity flagged: large transaction to an unknown wallet."
}

Reasoning Tree:
Step 1: Analyze the transaction value.
    - Branch 1: The value (500 ETH) is exceptionally high.
    - Branch 2: The value might be legitimate if the sender is a known whale.

Step 2: Evaluate the receiver address.
    - Branch 1: The receiver address is unknown, raising suspicion.
    - Branch 2: The receiver might be new or recently created, requiring further verification.

Step 3: Assess transaction context (notes, history).
    - Branch 1: Notes indicate unusual activity, supporting suspicion of fraud.
    - Branch 2: No historical record of suspicious activity from the sender mitigates concerns.

Evaluation:
- Combining high transaction value, unknown receiver, and flagged notes, the likelihood of fraud is significant.

Decision: Fraudulent
Explanation: The transaction exhibits multiple risk factors, including a high value, an unknown receiver, and flagged notes, making it highly suspicious.

---

### Query ###

Transaction Data:
{
  "Transaction ID": "tx54321",
  "Sender Address": "0xABCDE12345F",
  "Receiver Address": "0xFEDCBA67890",
  "Value": "0.5 ETH",
  "Timestamp": "2024-11-26T10:45:00Z",
  "Notes": "Transaction from a personal wallet to another wallet."
}

Reasoning Tree:
Step 1: Analyze the transaction value.
    - Branch 1: ...
    - Branch 2: ...

Step 2: Evaluate the receiver address.
    - Branch 1: ...
    - Branch 2: ...

Step 3: Assess transaction context (notes, history).
    - Branch 1: ...
    - Branch 2: ...

Decision:
Explanation:
\end{prompt}
\end{tcolorbox}

\subsection{Graph-of-Thought (GoT)}
GoT prompting involves structuring the reasoning process as a graph, where nodes represent concepts, intermediate steps, or decisions, and edges capture relationships or dependencies between them.
\begin{tcolorbox}[prompt]
\begin{prompt}
### Instruction ###
You are an expert in blockchain analysis. For each query, build a graph of thoughts to reason through the relationships between transactions, addresses, and chains. Use the graph structure to explore dependencies and resolve the query systematically.

### Example 1 ###

Query:
"Is the transaction likely part of a cross-chain scam?"

Transaction Data:
{
  "Transaction ID": "tx12345",
  "Sender Address": "0xA1B2C3D4E5",
  "Receiver Address": "0x9876543210",
  "Chain": "Ethereum",
  "Value": "50 ETH",
  "Timestamp": "2024-11-25T14:30:00Z"
}

Related Transactions:
1. {
    "Transaction ID": "tx67890",
    "Sender Address": "0x9876543210",
    "Receiver Address": "bnb1ABC234DEF",
    "Chain": "Binance Smart Chain",
    "Value": "25 BNB",
    "Timestamp": "2024-11-25T14:40:00Z"
  }
2. {
    "Transaction ID": "tx54321",
    "Sender Address": "bnb1ABC234DEF",
    "Receiver Address": "0x112233445566",
    "Chain": "Ethereum",
    "Value": "20 ETH",
    "Timestamp": "2024-11-25T14:50:00Z"
  }

Graph Reasoning:
- Node 1 (Transaction tx12345):
  - Sender: 0xA1B2C3D4E5
  - Receiver: 0x9876543210
  - Value: 50 ETH
  - Chain: Ethereum
- Node 2 (Transaction tx67890):
  - Sender: 0x9876543210
  - Receiver: bnb1ABC234DEF
  - Value: 25 BNB
  - Chain: Binance Smart Chain
- Node 3 (Transaction tx54321):
  - Sender: bnb1ABC234DEF
  - Receiver: 0x112233445566
  - Value: 20 ETH
  - Chain: Ethereum

Relationships:
- Edge between Node 1 and Node 2:
  - The receiver in tx12345 (0x9876543210) is the sender in tx67890, suggesting fund transfer between Ethereum and Binance Smart Chain.
- Edge between Node 2 and Node 3:
  - The receiver in tx67890 (bnb1ABC234DEF) is the sender in tx54321, indicating a cross-chain transfer back to Ethereum.

Evaluation:
- The transactions form a triangular flow across Ethereum and Binance Smart Chain, with value inconsistencies suggesting potential obfuscation.
- This pattern is common in cross-chain scams involving laundering funds across multiple chains.

Decision: Likely part of a cross-chain scam.
Explanation: The graph reveals interconnected transactions with suspicious cross-chain fund movements and value inconsistencies.

---

### Query ###

Query:
"Is the following transaction part of a larger fraudulent scheme?"

Transaction Data:
{
  "Transaction ID": "tx99999",
  "Sender Address": "0xABCDEF123456",
  "Receiver Address": "0x654321FEDCBA",
  "Chain": "Ethereum",
  "Value": "100 ETH",
  "Timestamp": "2024-11-26T10:30:00Z"
}

Related Transactions:
1. ...
2. ...

Graph Reasoning:
- Node 1:
  - ...
- Node 2:
  - ...
- Node 3:
  - ...

Relationships:
- Edge between Node 1 and Node 2:
  - ...
- Edge between Node 2 and Node 3:
  - ...

Evaluation:
...

Decision:
Explanation:
\end{prompt}
\end{tcolorbox}

\end{document}